# Small but not least changes: The Art of Creating Disruptive Innovations

Running title: Minor Changes, Major Disruptions

Youwei He[a,b,*] and Jeong-Dong Lee[a,b]

[a]Technology Management, Economics, and Policy Program, Seoul National University, 1 Gwanak-ro, Gwanak-gu 08826, Seoul, South Korea; [b]Integrated Major in Smart City Global Convergence, Seoul National University, 1 Gwanak-ro, Gwanak-gu, Seoul 08826, South Korea.

*Corresponding author's e-mail address: heyouwei2020@snu.ac.kr*

## Abstract

In the ever-evolving landscape of technology, product innovation thrives on replacing outdated technologies with groundbreaking ones or through the ingenious recombination of existing technologies. Our study embarks on a revolutionary journey by genetically representing products, extracting their chromosomal data, and constructing a comprehensive phylogenetic network of automobiles. We delve deep into the technological features that shape innovation, pinpointing the ancestral roots of products and mapping out intricate product-family triangles. By leveraging the similarities within these triangles, we introduce a pioneering 'Product Disruption Index'—inspired by the CD index (Funk and Owen-Smith, 2017)—to quantify a product's disruptiveness. Our approach is rigorously validated against the scientifically recognized trend of decreasing disruptiveness over time (Park et al., 2023) and through compelling case studies. Our statistical analysis reveals a fascinating insight: disruptive product innovations often stem from minor, yet crucial, modifications.

*Keywords:* product phylogenetic network, technological importance, product-family triangle, disruptive innovation, product disruption index (PDI), disruptive ancestral effect

# 1. Introduction

Innovation has a significant impact on the development of the economy and provides sustainable competitive advantage in management (Damanpour and Wischnevsky, 2006; Nagano et al., 2014). The disruptive innovation theory proposed by Christensen in 'The Innovator's Dilemma' more than 20 years ago has been widely discussed and applied (Christensen et al., 2018). In this theory (Christensen and Clayton M., 2013), the term 'disruptive technology' refers to technology that is inferior to mainstream technology valued by mainstream consumers but focuses on some neglected attributes instead. As this technology improves over time, it gradually surpasses the dominant technology in a given market. Specifically, a disruptive technology is hardly used in the early days but is used significantly in the later period. However, the definition of disruptive technological innovation within companies has remained ambiguous and indistinct (Roblek et al., 2021). In a typology proposed by Godart and Pistilli (2024), disruption can be categorized into technological (e.g., Sull, 1999; Tripsas and Gavetti, 2000; Danneels et al., 2017) and non-technological (Pino et al., 2016). Yet, the identification of disruptive innovation is fundamentally market-based. By discerning market forces, relative market sizes, and an innovation's capacity to forge new markets, Linton (2002) developed a model grounded in the Bass formula to ascertain the disruptiveness of innovation at a market level of analysis, which was subsequently adopted by Schmidt and Druehl (2008) to pinpoint disruptive innovations. Although the evolution of technology is also deemed a critical determinant of disruptive innovation (Paap and Katz, 2004; Myers, 2002), these studies have predominantly concentrated on the marketplace without offering insights into individual organizations (Nagy et al., 2016). Furthermore, empirical studies that have substantiated the universality of disruptive innovation theory and elucidated the principles governing its formation are scarce (Ben-Slimane, Diridollou, and Hamadache, 2020). In certain empirical research, citation networks have been employed to assess disruptive innovation (Dotsika et al., 2017; Wang et al., 2024); however, they have not qualified the disruptiveness of each innovation. Moreover, research utilizing product information networks to evaluate disruptiveness of products has not been previously documented in the literature. Given these existing gaps, we propose a method based on a product network and consolidating and destabilizing (CD) index (Funk and Owen-Smith, 2017) to discern disruptive products. This study contributes to assisting enterprises in leveraging product information to identify disruptive products and analyze them. Our analysis of disruptive products has revealed design principles that enable enterprises to create products with greater disruptive potential, emphasizing that 'small but not least step' of improvements can significantly foster disruptiveness.

Generally, citations are one of the most critical, simple, standard, and objective indicators of scientific influence (Didegah and Thelwall, 2013; Yan et al., 2012). In bibliometrics, Wu et al. (2019) measured scientific and technological advances using the disruption (D) index which is a citation-based indicator derived from the CD index of Funk and Owen-Smith (2017). They only quantified disruptiveness and did not provide a clear definition of disruption. Their basic concept was that when a scientific paper that cites an article also cites most of the references to this article, the article strengthens its scientific field. However, when the opposite occurs, i.e., when future references to an article do not acknowledge its intellectual predecessors, the article is considered to disrupt its field. Consolidating and destabilizing innovations are two types of innovation that promote the evolution of products. Destabilising innovation is disruptive and can determine the future evolution direction of a product more than consolidating innovation. In the citation network of scientific papers, papers associated with Nobel Prizes have high D indices. The CD index has been used as an indicator of disruptiveness in paper and patent citation networks (Park et al., 2023).

From a market perspective, disruptive innovations are poised to capture the incumbent's market share (Christensen, 2006; Christensen & Raynor, 2003). Conversely, from a technological standpoint, latecomers are inclined to adopt technologies from disruptive product rather than only following the incumbents' technology trajectory, potentially giving rise to new ecosystems and disrupting established industries (Kumaraswamy et al., 2018; Palmié et al., 2020; Ozalp et al., 2018; Silva & Grützmann, 2023; Dedehayir et al., 2014; Dedehayir et al., 2017). Consequently, the design of disruptive products is of paramount importance for businesses (King and Baatartogtokh, 2015). However, unlike papers or patents, there are no citations between products; instead, there are similarities between products. In this study, we defined a product disruption index (PDI) as an index that can measure the extent of product disruptiveness. First, we exploited the similarities between products to construct a phylogenetic network based on the focal product. The PDI was then calculated using the CD index in the phylogenetic network. The PDI was validated by several approaches. Park et al. (2023) observed that over time, the rate of disruptive innovation in the scientific and technological fields has decreased. Drawing a parallel to the product innovation domain, it is deduced that early iterations within a product series often exhibit a greater potential for disruption. This trend can be attributed to the rapid growth rate of the number of consolidated products outpacing that of disruptive products. Firstly, we verified the PDI for all car series and found that, on average, earlier models within the series are more disruptive, which is consistent with previous inferences. Secondly, we conducted a case study to further validate the PDI by comparing models from Tesla and Chevrolet. This case study revealed that Tesla's product, despite being a smaller change from its ancestor, still holds a significant PDI. In contrast, Chevrolet's electric vehicles (EVs) have undergone considerable changes from their predecessors but exhibit a lower PDI. This underscores the principle of 'small but not least' (SBNL) significantly promote disruptiveness. Thirdly, we performed another case study comparing regular and luxury vehicles. Our findings indicate that luxury brands command higher prices and exhibit less disruption compared to regular brands. This observation is consistent with the theory of disruptive innovation, which posits that from a market perspective, disruptive innovation often utilizes more affordable, simpler, and user-friendly technology to capture lower-end or emerging market share and challenge established technologies (Christensen, 2000; Christensen and Raynor, 2003; Govindarajan and Kopalle, 2006; Tellis, 2006).

This research focused on product technology and analyses the relationship between the creation of disruptive products and the technological changes involved. Most disruptive innovation analyses are conducted from the perspective of a market or business model (Benzidia et al. 2021; Christensen and Raynor 2013). However, the design principle of disruptive products of 'small but not least changes' is technological. To support this argument, we reviewed the relevant literature on disruptive innovation. Govindarajan and Kopalle (2006) proposed that radicalness and disruptiveness are distinct concepts. Radicalness is related to disruptive innovation but is not the only factor. Adner (2002) classified the attributes of disruptive innovation into two types: inherited and novel attributes. He argued that the competition between disruptive innovation and existing technology occurs in these dimensions. Yu and Hang (2011) suggested the following research and development strategies for disruptive innovation miniaturisation, simplification, augmentation, and exploitation. The above studies provide theoretical insights and practical guidance for the creation of disruptive products. Section 1 introduced disruptive innovation theory and objective for measuring the disruptiveness of products. Section 2 introduces relevant research on the product phylogenetic networks (O'Brien et al., 2001; Khanafiah and Situngkir, 2006; Tëmkin and Eldredge, 2007) from the product evolution theory (Yoon et al., 2024; Lee et al., 2021; Tellis et al., 1981; Massey, 1999; Utterback and Abernathy, 1975) and CD index from the bibliometrics field which are the major methods for calculating product disruptiveness. Section 3 introduced the framework and methods of how to construct a product phylogenetic network for automobile data and presents the PDI calculation. It also includes the validation of the PDI and two case studies. Section 4 discusses the regression analysis. Section 5 presents the conclusions and limitations of the study.

## 2. Literature review

2.1 Disruptive innovation

As a forefront issue in innovation and strategic management research, incumbents often fail to pursue new avenues of growth and are incrementally surpassed by new entrants, a phenomenon defined as disruptive innovation (Christensen, 1997). Disruptive innovation, defined at the market level as a process (Levina, 2017), lends itself to qualitative analysis but presents challenges for quantitative research. Schmidt and van der Sijde (2022) classified disruptive business models into categories such as Matchmakers, Standardizers, Service Providers, Open Collaborators, and Performance Reducers, which facilitates a taxonomy of disruptive business models but does not allow for quantitative study. Quantitative research and forecasting of disruptive innovation are notably difficult (Linton, 2002). In contrast, disruptive products, as man-made artifacts distinct from business models and not part of a disruptive process, should have quantifiable disruptiveness. However, current strategic management theories on disruptive innovation lack a method for quantifying the disruptiveness of products. Additionally, from a technological perspective, existing theories do not adequately explain the relationship between disruptive products and their ancestors. While some scholars have attempted to interpret disruptive innovation through technology diffusion—Chen et al. (2016) posited that technical performance trajectories, including disruptive technologies, follow an S-curve, and Linton (2002) attempted to forecast market diffusion of disruptive and discontinuous innovation using Bass's model (1969)—Wang et al. (2024) employed patent citation network maximum likelihood fitting methods and goodness-of-fit tests to analyze the network's degree distribution characteristics throughout its evolution. Yet, there has been no measurement of the disruptiveness of products.

2.2 Technology recombination and imitation strategy

Usher (1954) defines technological invention as the constructive assimilation of pre-existing elements into a new synthesis. Subsequently, scholars in economics and innovation management have recognized that technological novelty is driven by the recombination of existing knowledge and technology (e.g., Nelson and Winter, 1982; Weitzman, 1998; Arthur, 2009). Arts and Veugelers (2015) argue that characterizing technological invention as an evolutionary and recombinant search process, where the creation of new inventions by combining formerly disparate technology components, is a key process leading to more useful inventions.

Through an empirical analysis of the early U.S. automobile industry, Argyres et al. (2015) argue that innovation shocks can lead to significant changes in industry dynamics. This creates a follower's dilemma for other competitors, which is whether to imitate an innovation introduced by a competitor. According to Lee et al. (2012), learning from previous products and recombining features to create new ones can be categorized into imitation (directly copying successful products with little modification) and creative adaptation (learning from successful products and making some improvements). They conclude that creative adaptation has a stronger positive effect on financial performance (e.g., return on assets) compared to pure imitation. Additionally, Liao (2022), through analyzing the Small and Medium-sized Enterprises (SMEs), suggests from the perspective of 'learning by doing' that firms can improve their skills and capabilities through the process of imitating existing products, providing empirical evidence that adaptation leads to greater skill improvement compared to imitation. Furthermore, Dell'Era and Verganti (2007) propose that an imitation strategy involves companies quickly adopting new product attributes that appear in the market, usually with higher product attribute heterogeneity, as they tend to experiment with multiple product attribute combinations. While these studies highlight that imitation can promote innovation, they do not address how to select the objects of imitation and extract knowledge from them to develop disruptive products. This paper fills this theoretical gap and proposes the principle of "small but not least changes" in product design, to help designers create more disruptive products.

## 2.3 Product evolution and phylogenetic networks

A phylogenetic network is a method used in biology to analyze the evolution of species. A product can be viewed as a combination of technologies, and various technologies can be regarded as the genes of products (Arthur, 2007). The evolution of artificial products also mirrors biological evolution, adhering to the principle of survival of the fittest in the marketplace, and continuously learning and evolving through mutual influence (Yoon et al., 2024; Lee et al., 2021; Tellis et al., 1981; Massey, 1999; Utterback and Abernathy, 1975). With the technological dissemination, products show similarities over time, forming a product lineage. This lineage between products is not limited to the same brand or company because technology genes can spread across brands or categories. Phylogenetic networks (O'Brienet al., 2001; Khanafiah and Situngkir, 2006; Tëmkin and Eldredge, 2007) can be constructed based on the similarity between products. Lee et al. (2022) introduced a novel method for constructing phylogenetic product networks. Phylogenetic networks in biology and products are distinctly different, requiring speculating on the sequence in which species evolved. In the realm of product evolution, phylogenetic networks are employed to examine the technology life cycle, product lineages, and technology dynamics, etc. (Jeong et al., 2023; Lee et al., 2023; Jeong and Lee, 2024; Park et al., 2024). In previous literature, Lee et al. (2022) utilized phylogenetic network analysis on smartphone product data to explain the evolutionary process of smartphones and predict future product types. The phylogenetic network has also been applied to power sector institution data, proposing that both institutional inertia and ecological pressure influence the dynamics of institutional evolution to be either gradual or rapid, respectively (Lee et al., 2023). Additionally, by constructing a phylogenetic tree for photovoltaic technology, Park et al. (2024) have demonstrated that diversity is essential for the evolutionary mechanism to function, and technology integration is the correct path to follow. Furthermore, through the construction of a phylogenetic network for mobile products, Jeong and Lee (2024) have introduced the concept of product lineage life cycle and indicated that the key to revival lies in maintaining the niche market of feature phones by preserving their specialty and gradually enhancing their innovativeness within it. Moreover, phylogenetic networks have been utilized in paper data (Jeong et al., 2023) to explore the evolutionary patterns of financial AI technology. The phylogenetic network framework offers a systematic approach to convert product information into chronologically sequenced product citation networks, utilizing the principles of product evolution theory.

2.4 Consolidating and Destabilizing

Funk and Owen-Smith (2017) proposed a novel index for measuring whether an invention consolidates or destabilises (CD) in the existing technology development path. This study was based on the foundational theory of technological change and innovation by Schumpeter's (1976). By conducting a study of patent and paper data, Park et al. (2023) found that the average CD index of papers and patents decrease over time. The CD index represents the disruptiveness of an innovation (Park et al.,2023). Technological innovations can be divided into two categories. The first one is innovation whose contribution improves existing knowledge. The second is innovation that can disrupt existing knowledge, make it obsolete, and propel the technological development path in new directions. The concept of the CD index is to characterise the consolidating or disruptive nature of science and technology. When a paper or patent is disruptive, the subsequent work that cites it is less likely to cite its ancestors appearing in references. However, if a paper or patent is consolidating type, a subsequent article that cites it is more likely to cite its ancestors. Based on the network formed by patent and paper citations, the CD index can be calculated as suggested by Funk and Owen-Smith (2017).

$$CD_t = \frac{1}{n_t}\sum_{i=1}^{n} -2f_{it}b_{it} + f_{it} \quad (1)$$

$f_{it} = 1$ if $i$ cites the focal paper/patent; 0 if not.
$b_{it} = 1$ if $i$ cites the ancestors of the focal paper/patent; 0 if not.
$n_t$: number of forward citations of the focal work and/or ancestors at time $t$.

Wu et al. (2019) simplified the CD index to evaluate the disruptiveness of academic papers, as expressed in equation (2). Although equation (2) differs from equation (1) in terms of form, their main concepts are the same. Instead of consolidating and destabilising, they use the terms developing and disruption. In this definition, an ancestor refers to a node cited by the focal node. The nodes that follow the focal node are classified into three types. First, type $i$ nodes are the nodes that cite the focal node but not any of the ancestors of the focal node. Second, type $j$ nodes are nodes that cite not only the focal node but also any of the ancestors of the focal node. Third, type $k$ nodes are nodes that cite an ancestor of the focal node, but not the focal node. In equation (2), the D index is calculated using the counts of the three types of nodes. When D < 0, the node is developing type. When $D$ is greater than zero, the node is considered disruptive. When D = 0, the node is neutral.

$$D = \frac{n_i - n_j}{n_i + n_j + n_k} \qquad (2)$$

$n_i$: the number of $i$ type nodes which cite the focal paper but not the ancestors of the focal node.

$n_j$: the number of $j$ type nodes which cite both focal node and any of its ancestor.

$n_k$: the number of $k$ type nodes which cite an ancestor of the focal node but not the focal node.

Wu et al. (2019) utilized the CD index method to analyze patent and paper data, resulting in the finding that large teams develop while small teams disrupt science and technology. Park et al. (2023) analyzed patent and paper data using the CD index, concluding that papers and patents are becoming less disruptive over time. Lin et al., (2023) analyzed patent and paper data using the CD index, leading to the conclusion that remote collaboration fuses fewer breakthrough ideas. Although Funk and Owen-Smith (2017) mention that the CD index is not limited to patent and paper citation networks, there has been no precedent in previous literature for using this method to measure the disruptiveness of products which is main because there is no explicit citation network in the product space. From equation (2), it is evident that the calculation of disruptiveness is determined by the quantity of various types of descendants associated with the focal node. This calculation is based on a visible citation network. In the product space, there is no such visible citation network, rendering the CD index method inapplicable directly. However, the CD index offers a method for measuring disruptiveness through a citation network that accounts for the chronological sequence of innovations.

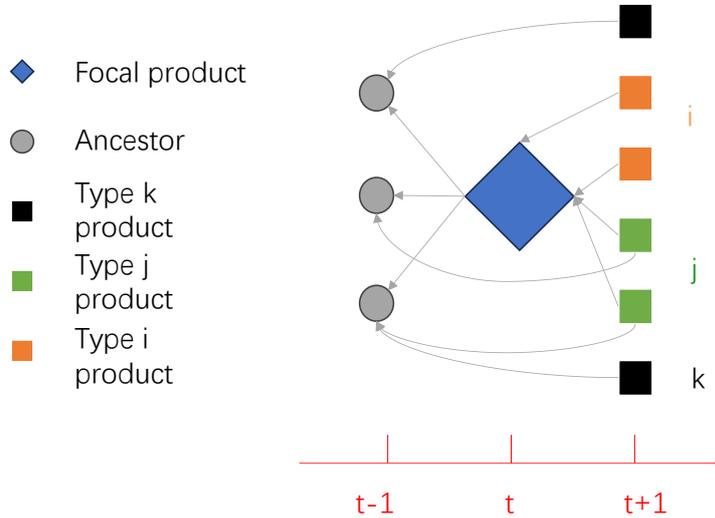

Figure 1. Part of product phylogenetic network. This network contains focal product in center. At time $t + 1$, type $i, j$, and $k$ products are present.

## 3. Methods

### 3.1 Data

We conducted an empirical case study using product data from the automotive industry. We collected car data from https://www.edmunds.com/. This website contains data on passenger car products sold in the US market. For each model of car, we obtained comprehensive evaluation scores from consumers through a website survey. The model years for the cars ranged from 2013 to 2024. This range represents the year of the car listing from 2012 to 2023 because according to the model-naming rules of a car are generally labelled with the next year. After cleaning the data, 4496 car models were created from 2013 to 2024. These car models involved 38127 technologies. Some of these technologies had continuous variables, such as the cruising range, whereas others had binary variables, such as engine type. However, we converted all technologies to the interval [0,1] by $l1$ normalization using equation (3). The $l1$ normalization (Albon and Chris, 2018) is a normalization technique that divides the gene values of each product in the dataset by the sum of the absolute values of that gene across all the products. This ensured that for each gene, the sum of the gene values for all products was 1. Our product data comprised a matrix with 38127 columns and 4496 rows. The columns of the matrix represented the technologies defined as product genes. A chromosome is a vector composed of all the genes in a product that can represent a unique product as a combination of technology genes. The rows of the matrix represent the products. This matrix is known as the chromosome matrix of products. However, when we calculated the one-year PDI, because the products in 2013 did not have ancestors, the data analysis started from the products in 2014. For the same reason, the data for 2024 could not be used. A total of 4114 products were produced between 2014 and 2023.

$$\hat{x}_{i,j} = \frac{x_{i,j}}{\sum_{i=1}^{n} |x_{i,j}|} \quad (3)$$

$x_{i,j}$: value of the $j$th gene of the $i$th product.
$\hat{x}_{i,j}$: normalized value of the $j$th gene of the $i$th product.

### 3.2 Conceptual framework

The constructing of a phylogenetic network required several steps as shown in Figure 2. First, the products must be represented as chromosomes. Second, a product similarity adjacency matrix is calculated using the chromosome and gene weight matrices. Third, a product phylogenetic network is constructed using the product similarity adjacency matrix. Fourth, the PDI is detected by the product phylogenetic networks based on each focal product. Finally, statistical analysis of all the variables is performed. The lost gene rate, inherited gene rate, and other variables are measured using both the product similarity adjacency matrix and gene weight matrix. Finally, statistical analysis of all the variables is performed. The lost gene rate refers to the rate at which gene values appear in ancestor genes but are not passed on to the focal product. The new gene rate refers to the ratio of gene values that appear in the focal product but not in the ancestors. The inherited gene rate is the ratio of the gene values inherited from the ancestors. Mathematical definitions of the parameters are provided in the following sections.

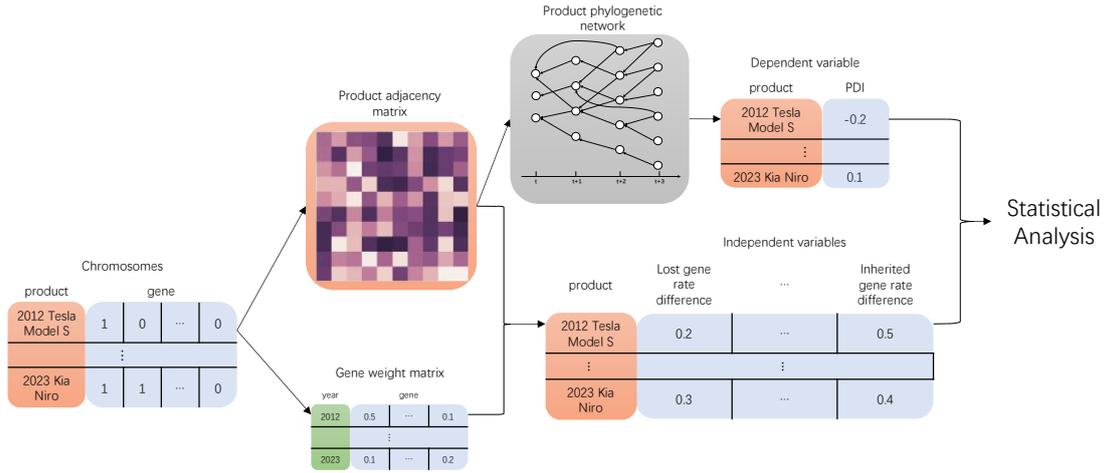

Figure 2. Conceptual framework

### 3.3 Methodology

The CD index is based on the number of citations; however, in the product space, there are similarities between products and not citations. The CD index, which is traditionally based on citation network, cannot directly apply to the product space due to the inherent similarities between products rather than citations. Consequently, we have enhanced the CD index to be compatible with a product phylogenetic network. In essence, we have redefined product similarities as citations and incorporated disruptiveness into equation (2) to compute the PDI. In the product space, the weight of an edge is the cosine similarity between products, and its value range is [0,1]. We converted the implicit similarities between products into explicit citations using a threshold. We deleted the edges whose similarity was less than the threshold and retained the edges whose similarity was greater than the threshold. A preserved edge was considered a citation.

Each product was a node in the product space. Ancestors were taken as the nodes in the years before the year of the focal node, and descendants were the nodes in the years after the year of the focal node. Figure 3 shows the product similarity space of a focal product, its ancestors, and descendants, and the product family triangle. In the product similarity space, the products are fully connected, and the link weight is the similarity between each pair of products. When we converted the product similarity space into an edge of the product phylogenetic network, we only needed to convert the links into citations using a threshold, which is a major method when construct a phylogenetic network (Jeong and Lee, 2024; Jeong et al., 2023; Park et al., 2024). The threshold for the product family triangle was obtained using equation (4). In Figure 3, it also shows a product-family triangle representing the relationship between a pair of an ancestor and a descendant associated with a focal product. In this triangle, the link between ancestor $k$ and focal node $j$, $a_{k,j}$ exists when the value is 1 and does not exist when the value is 0. Similarly, $f_{i,j}$ and $b_{k,i}$ are used to represent the links between descendant $i$ and focal node $j$, and the link between descendant $i$ and ancestor $k$. In equation (4), the threshold obtained is the minimum of two averages. The first average is the average of the cosine similarities between the focal node and its descendant and between the focal node and its ancestor in the product family triangle. The second is the average cosine similarity between the focal node and its descendants and between the focal node and its ancestors across all product-family triangles. In this study, using equations (5)-(7), we detected the citations of the focal node and its ancestors and citations of the focal node and its descendants. When the similarities were converted into citations, the product similarity space was converted into a product phylogenetic network.

$$threshold_{k,j,i} = \min\left(\frac{sim_{k,j}+sim_{i,j}}{2}, \sum_{i=1}^{n}\sum_{k=1}^{m}\frac{sim_{k,j}+sim_{i,j}}{2*m*n}\right) \quad (4)$$

$m$: the count of ancestors

$n$: the count of descendants

$k$: the $k$-th ancestor

$j$: the focal product

$i$: the $j$-th descendant

$sim_{k,j}$: cosine similarity between nodes $k$ and $j$

$$a_{k,j} = \begin{cases} 1 & sim_{k,j} > threshold \\ 0 & sim_{k,j} \leq threshold \end{cases} \quad (5)$$

$$f_{i,j} = \begin{cases} 1 & sim_{i,j} > threshold \\ 0 & sim_{i,j} \leq threshold \end{cases} \quad (6)$$

$$b_{k,i} = \begin{cases} 1 & sim_{k,i} > threshold \\ 0 & sim_{k,i} \leq threshold \end{cases} \quad (7)$$

$a_{k,j}$: the citation from the focal product and the $k$-th ancestor; 1: with citation and 0: no citation.

$f_{i,j}$: citation between the $i$-th descendant of the focal product and the focal product; 1: with citation and 0: no citation.

$b_{k,i}$: citation from the $i$-th descendant of the focal product and $k$-th ancestor; 1: with citation, 0: no citation.

$threshold_{k,j,i}$: the threshold in the product-family triangle formed by nodes $i, j, k$.

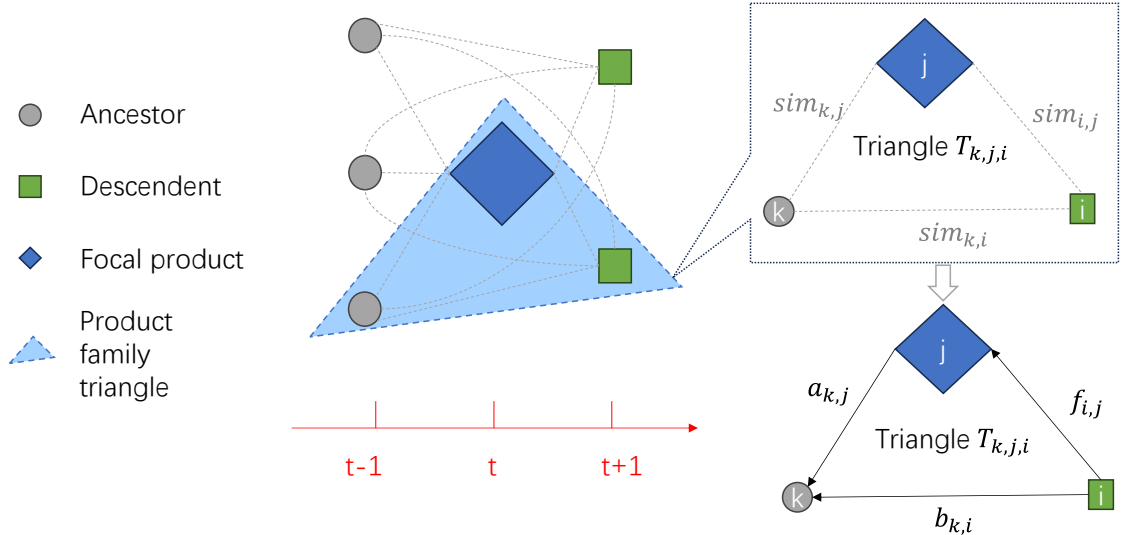

Figure 3. Product similarity space, including focal product, its ancestors, and descendants, and product-family triangle. In product-family triangle, we can transfer similarities between products into citations. $sim_{k,j}$ is similarity between product $k$ and $j$, $sim_{i,j}$ is similarity between product $i$ and $j$, and $sim_{k,i}$ is similarity between product $k$ and $i$. $a_{k,j}$ is citation from node $j$ to node $k$, $f_{i,j}$ is citation from node $i$ to node $j$, and $b_{k,i}$ is the citation from node $i$ to node $k$.

If $p$ is the count of focal products, $n$ is the count of ancestors and $m$ is the count of descendant products, there are $m*n$ product family triangles for one focal product. Because there is only one focal product, $p$ should equal 1. We obtained the values of $a_{k,j}$, $b_{k,i}$, and $f_{i,j}$ using equations (8)-(10). When we obtained all the citations, we constructed a phylogenetic network centered around the focal node, spanning generations over a year. Using all the triangles centred on focal node $i$, we detected the type of focal node $i$ using equations (8)-(10). Equations (8)-(10) are the conditions for determining whether the descendant nodes belong to types $i, j$, and $k$, respectively. To prevent the same point from being determined as a different type, we stipulated that the order of the considered points was $j, i, k$. After detecting the number of descendants of type $i, j$, and $k$ of the focal node, we used equation (2) to calculate the PDI of the focal product.

$$\sum_{k \in anc, i \in des} (a_{k,j} = 1 \land b_{k,i} = 1 \land f_{i,j} = 1) > 0$$
$$\Rightarrow node_i \in \{x \in X | type(x) = 'j'\} \quad (8)$$

$$(\sum_{k \in anc, i \in des} (b_{k,i} = 0 \land f_{i,j} = 1) > 0) \land (node_i \notin \{x \in X | type(x) = 'j'\})$$
$$\Rightarrow node_i \in \{x \in X | type(x) = 'i'\} \quad (9)$$

$$(\sum_{k \in anc, i \in des} (a_{k,j} = 1 \land b_{k,i} = 1 \land f_{i,j} = 0) > 0) \land (node_i \notin \{x \in X | type(x) \in \{'j', 'i'\}\})$$
$$\Rightarrow node_i \in \{x \in X | type(x) = 'k'\} \quad (10)$$

$node_i$: $i$-th node of descendants of the focal node.
$type(x)$: Function used to obtain the type of node, which is a node following the focal node.
$anc$: Node set of the ancestors of the focal node.
$des$: Node set of descendants of the focal node.

3.4 Product genetic variables

Technological importance can be represented by the weight of a gene. The weight of a gene can be determined through various methods. For instance, Lee et al. (2023) utilized normalized entropy weight, while Lee et al. (2022) and Jeong et al. (2023) applied the Term Frequency – Inverse Document Frequency (TF–IDF) method to calculate gene weight. In this study, the weight of a gene increases with the number of products it is found in. Therefore, we used frequency weight in equation (11) to calculate the gene's weight.

$$W_i^T = \frac{\sum_{j=1}^{n} x_{i,j}}{n} \quad (11)$$
$$W_i^T \in [0,1]$$

$W_i^T$: $i$th gene weight in year $T$.
$x_{i,j}$: $i$th gene value of the $j$th product.
$n$: product count in year $T$.

From a microscopic and evolutionary standpoint, we explore the role of a technology gene as a catalyst for technological advancements. The combination of technologies to create a product is driven by three distinct search patterns (Park et al., 2024): vertical inheritance (VI), as described by Lawrence (2005); horizontal gene transfer (HGT), as studied by Carignani et al. (2019); and mutation (MT), as outlined by Kardong (2008).

During the process of Vertical Inheritance (VI), descendants inherit a wealth of genetic information from their direct ancestors (Lawrence, 2005). Existing ancestors serve as the benchmark for evolution, providing the foundational structure of genetic composition (Carignani et al., 2019; Wagner and Rosen, 2014). Descendants then engage in a gradual evolutionary process, guided by the genetic traits passed down from their forebears (Anderson and Tushman, 1990; Tellis and Crawford, 1981). The inherited gene rate, $inhr_j$, is the rate of the summation of all the inherited gene values in the focal product to the sum of all the gene values in the focal product. The weighted inherited gene rate, $inhwr_j$, is the summation of all the weighted inherited gene values in the focal product to the summation of all the weighted gene values in the focal product. The inherited gene rate difference, $inhr\_diff_j$, is the difference between the weighted inherited gene rate and the inherited gene rate. This represents the degree of genetic importance inherited by a product from its ancestors. The equations (12)-(14) provide the corresponding mathematical definitions. The inherited gene rate measures the degree of VI, while the weighted inherited gene rate quantifies the extent of VI for significant genes. We can determine whether a product has inherited more crucial genes or less important ones by examining the magnitude of the inherited gene rate difference.

$$inhr_j = \frac{\overline{\sum_{i}^{i \in anc(j) \cap foc(j)} x_{i,j}}}{\sum_{i}^{i \in foc(j)} x_{i,j}} \quad (12)$$

$$inhwr_j = \frac{\overline{\sum_{i}^{i \in anc(j) \cap foc(j)} x_{i,j} \cdot w_i^T}}{\sum_{i}^{i \in foc(j)} x_{i,j} \cdot w_i^T} \quad (13)$$

$$inhr\_diff_j = inhwr_j - inhr_j \quad (14)$$

$inhr_j$: average inherited gene rate of the $j$th product, $inhr_j \in [0,1]$.
$inhwr_j$: average weighted inherited gene rate of the $j$th product, $inhr_j \in [0,1]$.
$inhr\_diff_j$: inherited gene rate difference of the $j$th product.
$x_{i,j}$: $i$th gene value of the $j$th product.
$anc(j)$: function for obtaining all the genes of the ancestors of the $j$th product.
$foc(j)$: function for obtaining all the genes of the $j$th product.
$w_i^T$: $i$th gene weight in year $T$.

The lost gene rate can be utilized to measure the rate of gene loss during the VI process. The weighted lost gene rate assesses the extent of loss for significant genes. By examining the lost gene rate difference, we can ascertain whether more crucial genes or less important ones have been lost. The lost gene rate, $lostr_j$, is the rate of summation of all the lost gene values in the focal product to the summation of all the gene values in the ancestors. The weighted lost gene rate, $lostwr_j$, is the rate of summation of all the weighted lost gene values in the focal product to the summation of all the weighted gene values in the ancestors. The lost gene rate difference, $lostr\_diff_j$, is the difference between the weighted lost gene rate and the actual lost gene rate. This represents the degree of genetic importance of a product lost from its ancestor.

$$lostr_j = \frac{\overline{\sum_{i}^{i \in anc(j) - foc(j)} x_{i,j}}}{\sum_{i}^{i \in anc(j)} x_{i,j}} \quad (15)$$

$$lostwr_j = \frac{\overline{\sum_{\iota}^{\iota \in anc(j)-foc(j)} x_{\iota,J} \cdot w_\iota^T}}{\sum_{\iota}^{\iota \in anc(j)} x_{\iota,J} \cdot w_\iota^T} \quad (16)$$

$$lostr\_diff_j = lostwr_j - lostr_j \quad (17)$$

$lostr_j$: average lost gene rate the $j$th product, $lostr_j \in [0,1]$.
$lostwr_j$: average weighted lost gene rate of the $j$th product, $lostwr_j \in [0,1]$.
$lostr\_diff_j$: gene lost rate difference of the $j$th product.

During the process of HGT which is primarily driven by asexual reproduction in biology, descendants acquire new genetic information from a neighboring ancestor of a different lineage, rather than from a direct ancestor (Lawrence, 2005; Smets and Barkay, 2005). In the process of Mutation (MT), new traits emerge in the descendant that were not present in previous generations. This mechanism facilitates the emergence of new genetic elements and their combination to form novel genetic configurations (Kardong, 2008). The new gene rate is a metric used to measure the proportion of new genes acquired during MT and HGT processes. The weighted new gene rate assesses the proportion of significant new genes, while the new gene rate difference indicates whether the product has obtained more important or less important new genes. The new gene rate, $newr_j$, is the rate of the summation of all the new gene values in the focal product to the summation of all the gene values in the focal product. The weighted new gene rate, $newwr_j$, is the rate of the summation of all the weighted new gene values in the focal product to the summation of all the weighted gene values in the focal product. The lost gene rate difference, $newr\_diff_j$, is the difference between the weighted new gene rate and the new gene rate. This represents the degree of genetic importance of the new genes of a product.

$$newr_j = \frac{\overline{\sum_{\iota}^{\iota \in anc(j)-foc(j)} x_{\iota,J}}}{\sum_{\iota}^{\iota \in foc(j)} x_{\iota,J}} \quad (18)$$

$$newwr_j = \frac{\overline{\sum_{\iota}^{\iota \in anc(j)-foc(j)} x_{\iota,J} \cdot w_\iota^T}}{\sum_{\iota}^{\iota \in foc(j)} x_{\iota,J} \cdot w_\iota^T} \quad (19)$$

$$newr\_diff_j = newwr_j - newr_j \quad (20)$$

$newr_j$: average new gene rate of the $j$th product, $lostr_j \in [0,1]$.
$newwr_j$: average weighted new gene rate of the $j$th product, $lostwr_j \in [0,1]$.
$newr\_diff_j$: new gene rate difference of the $j$th product.

The brand-new gene rate specifically measures the proportion of new genes obtained solely during the MT process. The brand-new gene rate, $ngr_j$, is the ratio of the brand-new gene count in the focal product to the gene count in the focal product.

$$X_t = \{x | x \text{ in the years earlier than } t\} \quad (21)$$

$$ngr_j = \frac{\sum_{x_{i,j} \notin X_t} x_{i,j} \cdot I(x_{i,j}>0)}{\sum x_{i,j} \cdot I(x_{i,j}>0)} \quad (22)$$

$X_t$: set of the genes in years earlier than year $t$.
$ngr_j$: brand-new gene rate of the $j$th product.
$I(x > 0)$: when $x > 0$, the value of the function is 1; otherwise, it is 0.

3.5 Descriptive Statistics and Correlations

We calculated the product adjacency similarity matrix with dimensions (4496 × 4496) using the chromosome matrix. It represents the similarity between pairs of products. Then, using the product adjacency similarity matrix, we built a product phylogenetic network and calculated the PDI for each product within one year using equation (2). The 1-year PDIs represent the short-term disruptiveness of a product. The average 1-year PDI was recorded at -0.296, suggesting that products with a one-year generational interval, on average, did not exhibit disruptiveness. Contrarily, the average CD index for patents and papers calculated by Park et al. (2023) was above zero, which means disruptive. This discrepancy can be attributed to the presence of 'multiples' in the technology field. 'Multiples' refer to a more than one of highly similar innovations developed independently (Merton, 1936; Merton, 1968; Wagner and Rosen, 2014). Consequently, in the explicit citation network, a focal node may lack citation links with closely related ancestors, leading to the omission of some ancestors—a not uncommon occurrence (Heneberg, 2013). Such omissions can result in a reduced count of j-type and k-type descendants, which, according to Equation (2), would lead to an overestimation of the focal node's CD index. However, this issue of omitted ancestors does not arise within the phylogenetic network, as our methodology involves calculating a citation network generated post-assessment of all ancestors' similarity to the focal product. Therefore, in comparison to the PDI, the CD index may overestimate the disruptiveness of the focal node. Figure 4.a shows the distribution of the 1-year PDI. The number of disruptive products is much smaller than the number of developing products. Figure 4.b shows the average 1-year PDI trend by year. A clear downward trend is observed from 2015 to 2023. The increase in the PDI in 2015 appears to be due to the large influx of electric vehicles into the market in 2015 and the emergence of various new technologies.

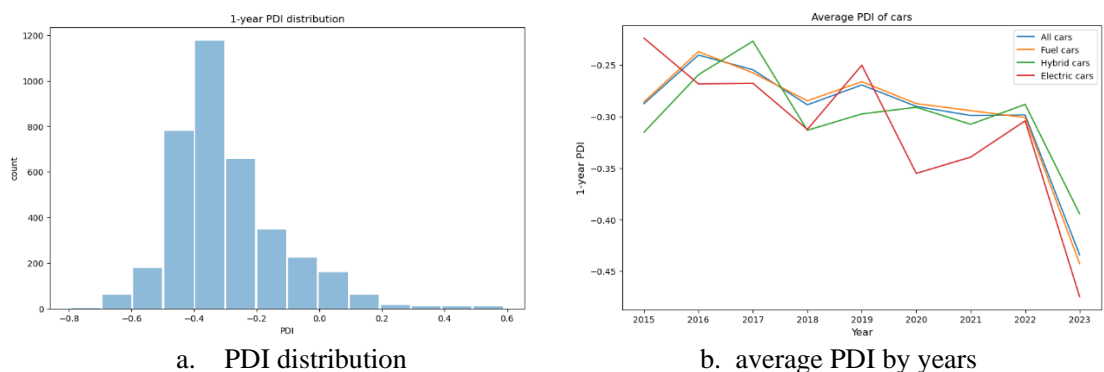

a. PDI distribution                          b. average PDI by years

Figure 4. One-year PDI distribution

Table 2 summarizes the descriptive statistics of the variables. The PDI is positively correlated with the lost gene rate difference, inherited gene rate difference, brand-new gene rate, and ancestor PDI. It is negatively correlated with the new gene rate difference. The ancestor PDI is the average PDI value of the ancestors of the focal product. We chose the three products that were most similar to the focal product and appeared earlier than the focal product as its ancestors. These relationships are presented in Table 2 and Figure 5. The PDI is positively related to the ancestral PDI. Therefore, disruptive products are likely to have disruptive ancestors. This effect may be caused by the one-year intergenerational phylogenetic network and one-year PDI.

Table 2. Descriptive Statistics and Correlations

| Variable | Mean | SD | Max | Min | 1 | 2 | 3 | 4 | 5 |
|---|---|---|---|---|---|---|---|---|---|
| 1. PDI | -0.296 | 0.173 | 0.589 | -0.795 | 1 | | | | |
| 2. Weighted lost gene rate | 0.061 | 0.052 | 0.543 | 0 | -0.073 | 1 | | | |
| 3. Unweighted lost gene rate | 0.197 | 0.122 | 0.853 | 0.009 | 0.004 | 0.83 | 1 | | |
| 4. Lost gene rate difference | -0.136 | 0.084 | 0.03 | -0.701 | -0.052 | -0.585 | -0.938 | 1 | |
| 5. Weighted inherited gene rate | 0.935 | 0.05 | 1 | 0.551 | 0.009 | -0.737 | -0.744 | 0.623 | 1 |
| 6. Unweighted inherited gene rate | 0.789 | 0.125 | 0.991 | 0.118 | -0.01 | -0.802 | -0.979 | 0.924 | 0.776 |
| 7. Inherited gene rate difference | 0.146 | 0.092 | 0.761 | -0.195 | 0.019 | 0.69 | 0.928 | -0.919 | -0.512 |
| 8. Weighted new gene rate | 0.065 | 0.05 | 0.449 | 0 | -0.009 | 0.737 | 0.744 | -0.623 | -1 |
| 9. Unweighted new gene rate | 0.211 | 0.125 | 0.882 | 0.009 | 0.01 | 0.802 | 0.979 | -0.924 | -0.776 |
| 10. New gene rate difference | -0.146 | 0.092 | 0.195 | -0.761 | -0.019 | -0.69 | -0.928 | 0.919 | 0.512 |
| 11. Price(log) | 10.74 | 0.687 | 14.164 | 9.392 | -0.039 | -0.009 | -0.014 | 0.015 | 0.005 |
| 12. Review rate | 4.624 | 0.362 | 5 | 1 | 0.051 | 0 | -0.01 | 0.014 | 0.009 |
| 13. Brand-new gene rate | 0.083 | 0.089 | 0.633 | 0 | 0.35 | 0.151 | 0.289 | -0.327 | -0.064 |
| 14. Ancestor PDI | -0.34 | 0.234 | 0.763 | -0.832 | 0.551 | -0.102 | 0.058 | -0.147 | 0.019 |

| Variable | 6 | 7 | 8 | 9 | 10 | 11 | 12 | 13 | 14 |
|---|---|---|---|---|---|---|---|---|---|
| 6. Unweighted inherited gene rate | 1 | | | | | | | | |
| 7. Inherited gene rate difference | -0.939 | 1 | | | | | | | |
| 8. Weighted new gene rate | -0.776 | 0.512 | 1 | | | | | | |
| 9. Unweighted new gene rate | -1 | 0.939 | 0.776 | 1 | | | | | |
| 10. New gene rate difference | 0.939 | -1 | -0.512 | -0.939 | 1 | | | | |
| 11. Price(log) | 0.014 | -0.016 | -0.005 | -0.014 | 0.016 | 1 | | | |
| 12. Review rate | 0.004 | -0.001 | -0.009 | -0.004 | 0.001 | -0.012 | 1 | | |
| 13. Brand-new gene rate | -0.279 | 0.344 | 0.064 | 0.279 | -0.344 | -0.07 | 0.009 | 1 | |
| 14. Ancestor PDI | -0.077 | 0.115 | -0.019 | 0.077 | -0.115 | 0.018 | 0.045 | 0.364 | 1 |

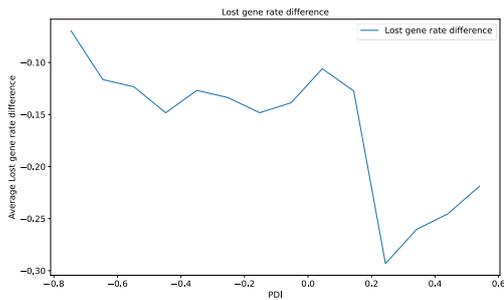

a. Average PDI plots of lost gene rate difference

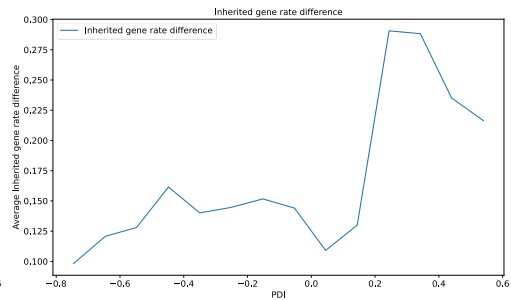

b. Average PDI plots of inherited gene rate difference

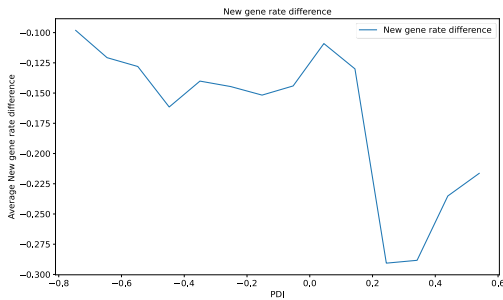

c. Average PDI plots of new gene rate difference

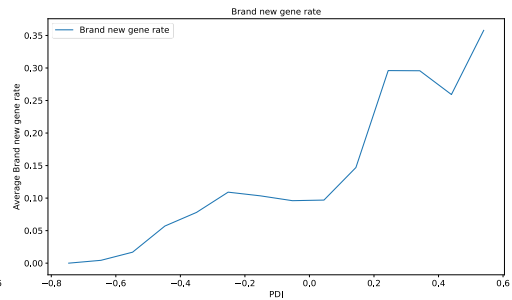

d. Average PDI plots of brand-new gene rate

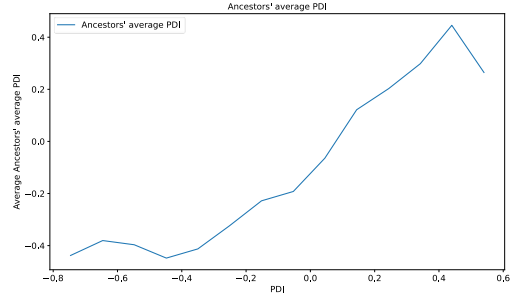

e.  Average PDI plots of ancestor PDI

Figure 5. Trends of independent variables with PDI

3.4 Validation with car series

Validation of the PDI was difficult because there were no real disruption data for the products. Even in the validation of the CD Index in previous literatures, its absolute correctness could not be guaranteed (Bornmann et al.,2020; Ruan et al.,2021). However, the PDI proposed in this study was based on the CD index, and others have validated on the CD index (Funk and Owen-Smith, 2017). Our validation of the PDI was grounded in the widely accepted notion that automobile models are innovations. In line with the research of Park et al. (2023), which demonstrates that technology and science are becoming less disruptive across various domains, it can be inferred that earlier models within a car series are more likely to be disruptive. Consequently, we applied this principle to validate our PDI results. The decreasing trend of the average PDI is illustrated in Figure 4.b. Subsequently, we will demonstrate the downward trajectory of PDI within car series.

For car products, among all the brands, many series exist, such as 'Model S' for Tesla. For products in the same series, the later products were expected to become less disruptive than the earlier products. Different series exhibited different PDI means and variances. Therefore, we used a normalized PDI in the series to compare the disruption of products within the series using equation (23). Owing to varying year counts across series, we employed the percentile in ascending order to denote the chronological order within each series. For example, if in a series, the products were from 2012, 2013, 2014, and 2015, the percentiles in ascending order for each product in this series were 0, 0.33, 0.66, and 1. We averaged all the series to obtain Figure 6. In Figure 6, the normalized PDI decreases as the time at which the product appears in the series increases. Specifically, the earlier a product appears in the series, the more disruptive it becomes. To a certain degree, this corroborates that the product disruptiveness outcomes derived from our methodology align with the findings of Park et al. (2023), which observed a decline in disruptiveness over time within paper and patent data.

$$y_{norm} = \frac{y - y_{min}}{y_{max} - y_{min}} \quad (23)$$

$y_{norm}$: normalized PDI of the product in series.
$y$: PDI of the product.
$y_{min}$: minimum PDI in the series.
$y_{max}$: maximum PDI in the series.

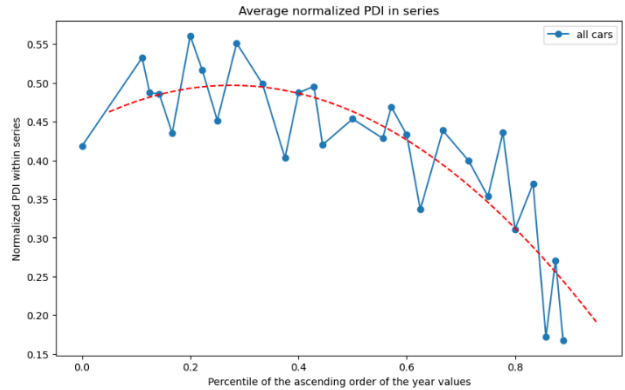
Figure 6. Average normalized PDI in series

3.5 A Case Study of Tesla and Chevrolet

As a case study, we compared the differences between Tesla Model S and the Chevrolet Spark EV in 2014. Both are early-stage electric vehicles. The 2014 Tesla Model S is the third generation of Model S and the 2014 Chevrolet Spark EV is the first generation of Spark EV. The details are presented in Table 3. The gene count is the total gene count of a product. The median of the weights of all the genes in 2014 was used as the threshold value. Genes with weights greater than this threshold were considered significant. The first ancestor refers to the product that is most similar to the product one year prior. The inherited genes were from the first ancestor. New genes were present in the product but not in the first ancestor. The review rate is the rate of the website surveys. This is the comprehensive score given by the consumers for each model. The score ranged from 0 to 5; the higher the score, the better the performance. Figure 7 shows the PDI positions of the two products in the PDI distribution.

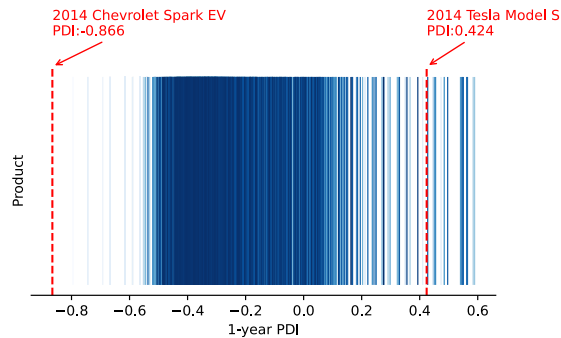
Figure 7. Positions of the two car models in the PDI distributions

In Table 3, the PDI of the 2014 Tesla Model S is positive and much higher than that of the 2014 Chevrolet Spark EV, consistent with the general perception of the review rate. Notably, Tesla is more similar to its ancestor and has inherited more genes. Chevrolet is very different from its ancestors and has adopted many new genes. Chevrolet seems to be developing new technologies; however, its product disruption rate is low. As a pioneer in electric vehicles, Tesla has maintained consistency in its design language compared to its predecessors, making only a few critical technological changes. In contrast, Chevrolet has modified more key elements from its predecessors. This aligns with the theory proposed by Dell'Era and Verganti (2007), which suggests that innovators tend to have less heterogeneity in their product language, while imitators exhibit more variations.

Table 3. Details of two electric vehicles

|  | 2014 Tesla Model S | 2014 Chevrolet Spark EV |
|---|---|---|
| Gene count | 147 | 152 |
| Weight threshold | 0.003 | |
| Important genes | 97 | 119 |
| Unimportant genes | 50 | 33 |
| New important genes | 7 | 86 |
| New unimportant genes | 0 | 2 |
| Inherited important genes | 90 | 33 |
| Inherited unimportant genes | 50 | 31 |
| First ancestor | 2013 Mercedes-Benz CL-Class | 2013 GMC Savana 3500 |
| Similarity | 0.918 | 0.310 |
| PDI | 0.424 | −0.866 |
| Review rate | 4.8 | 4.5 |
| Price | 69900 | 26685 |

3.6 A Case Study of Luxury and Regular Cars

The same company produces cars of different brands. We selected three pairs of luxury and regular brands from 3 companies. We measured the average values of the lost gene rate difference, inherited gene rate difference, new gene rate difference, and PDIs for six brands, as presented in Table 4. For each company, the regular group had a higher PDI, lower lost gene rate difference, higher inherited gene rate difference, and higher brand-new gene rate than the luxury group.

Table 4. Average values of Luxury cars and Regular cars

| Company | Toyota Motor Corporation | | Hyundai Motor Group | | Volkswagen AG | |
|---|---|---|---|---|---|---|
| Brands | Toyota | Lexus | Hyundai | Genesis | Volkswagen | Audi |
| Type | Regular | Luxury | Regular | Luxury | Regular | Luxury |
| PDI | −0.322 | −0.348 | −0.327 | −0.356 | −0.306 | −0.321 |
| Price | 32807.991 | 53901.930 | 27206.361 | 53927.407 | 28665.450 | 64392.913 |
| Review rate | 4.646 | 4.628 | 4.591 | 4.519 | 4.588 | 4.629 |
| Lost gene rate difference | −0.183 | −0.178 | −0.182 | −0.128 | −0.198 | −0.165 |
| Inherited gene rate difference | 0.160 | 0.155 | 0.149 | 0.112 | 0.161 | 0.149 |
| Brand-new gene rate | 0.087 | 0.078 | 0.087 | 0.036 | 0.110 | 0.080 |

Luxury brands have higher prices and less disruption than regular brands. Because regular brands have positive and higher inherited gene rate differences, they inherit more important genes, which account for the majority of the total genes. Regular brands have negative and lower lost gene rate differences indicating that they lose more important genes, which are only a small portion of the total genes. This indicates small and important changes related to the disruption. Moreover, regular brands also have a higher brand-new gene rate, i.e., they adopt many new technologies for innovation.

Figure 8a and b shows scatter of the sums of log prices and PDI grouped by series and brands, respectively. Both figures show a similar decreasing trend. As the total price increases, the sum of the PDI decreases. This suggests that they are more luxurious with less disruption. Luxury brands develop more in terms of quality, but not disruption, on average. In this case, luxury cars may spend more on quality than on disruption. Therefore, innovation of a product cannot be determined based on its price alone. Because these calculations are based on sums rather than averages, these figures reflect the overall values rather than individual values.

a. Grouped by series
b. Grouped by brands

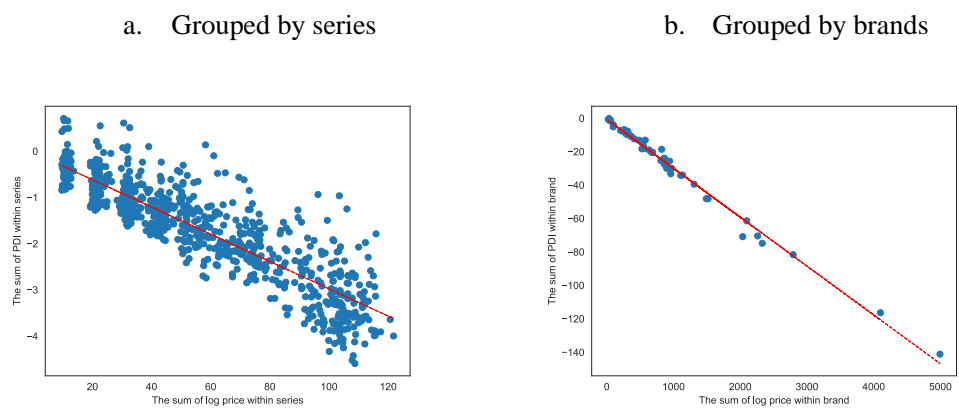

Figure 8. Scatters of the sum of grouped PDI and log price. X-axis represents sum of log price and Y-axis is sum of PDI.

3.7 Regression analysis

In the regression analysis, we used 1-year PDI as the dependent variable and the unweighted lost gene rate, unweighted inherited gene rate, unweighted new gene rate, lost gene rate difference, new gene rate difference, inherited gene rate difference, log product price, review rate from the website, and the average PDI of its three most similar ancestors as independent variables. Since not all car models receive a rating from the website, the analytical model incorporating the rating variable will consequently have a smaller number (3694) of observations compared to the original dataset (4496). Consequently, we performed six regression analyses, the details of which are summarized in Table 5. Given that each car model appeared only once, the dataset comprised cross-sectional data; hence, we employed pooled ordinary least squares (OLS) regression analysis. The PDI of the ancestors was determined by calculating the average PDI of the product's three most proximate predecessors. These predecessors are defined as the three products most similar to the focal product from all products released in the years preceding the focal product's launch. Our primary investigation centered on the influence of variations in critical and non-critical genes within the product's gene combinations on disruptiveness. Consequently, the explanatory variables included the inherited gene rate difference, the new gene rate difference, and the lost gene rate difference.

Table 5. Multiple regression results from years 2014 to 2023

|  | (1) | (2) | (3) | (4) | (5) | (6) |
|---|---|---|---|---|---|---|
| Unweighted lost gene rate | -0.038** (0.019) |  |  |  |  |  |
| Lost gene rate difference | | | | 0.064** (0.029) | | |
| Unweighted inherited gene rate | | | 0.045** (0.019) | | | |
| Inherited gene rate difference | | | | | | -0.087*** (0.026) |
| Unweighted new gene rate | | -0.045** (0.019) | | | | |
| New gene rate difference | | | | | 0.087*** (0.026) | |
| Price(log) | | | | -0.013*** (0.029) | -0.013*** (0.026) | -0.013*** (0.026) |
| Review rate | 0.013* (0.019) | 0.013* (0.019) | | | | |
| Ancestor PDI | 0.408*** (0.019) | 0.409*** (0.019) | 0.409*** (0.019) | 0.412*** (0.029) | 0.412*** (0.026) | 0.412*** (0.026) |
| const | -0.208*** (0.007) | -0.206*** (0.007) | -0.193*** (0.010) | -0.013 (0.004) | -0.008 (0.004) | -0.008 (0.004) |
| N | 3694 | 3694 | 3694 | 3694 | 3694 | 3694 |
| R2 | 0.306 | 0.306 | 0.305 | 0.307 | 0.309 | 0.309 |
| R2 Adj. | 0.305 | 0.305 | 0.305 | 0.307 | 0.308 | 0.308 |
| F-statistic | 541.197 | 541.967 | 810.49 | 546.019 | 549.039 | 549.039 |
| Log-Likelihood | 1911.598 | 1912.401 | 1910.538 | 1916.62 | 1919.758 | 1919.758 |
| VIF | 5.739 | 6.084 | 3.056 | 6.339 | 6.056 | 6.056 |

*Notes:* This table evaluates the relationship between different measures of the use of product genetic information, survey data from website, price and 1-year PDI. Estimates are from ordinary-least-squares regressions. Each coefficient is tested against the null hypothesis of being equal to 0 using a two-sided t-test. We do not adjust for multiple hypothesis testing. Robust standard errors are shown in parentheses.
+ p<0.1; * p<0.05; ** p<0.01; *** p<0.001.

Our models demonstrated an absence of multicollinearity, as indicated by the Variance Inflation Factor (VIF) values, all of which were below the threshold of 7. The PDI of ancestral products exhibited a positive correlation with the PDI of the focal product across all models, suggesting that products benefit from aligning with disruptive predecessors. The positive coefficient of the ancestral PDI provides compelling evidence of the disruptive ancestral effect. This suggests that in the realm of product evolution, assimilating insights from a disruptive predecessor can markedly increase the potential for innovation in the subsequent product. It echoes the proverbial wisdom that excellence begets excellence, akin to 'like father, like son.' In the context of the disruptive ancestral effect, the presence of a disruptive ancestor does not necessarily guarantee that the focal product will surpass its predecessor in disruptiveness. Rather, it indicates that the focal product is likely to exhibit a higher degree of disruptiveness compared to those without such an ancestor. This phenomenon underscores the advantage of 'standing on the shoulders of giants' to achieve greater disruptiveness. Additionally, the 'review rate' variable, derived from website review scores, showed a positive association with the PDI in models 1 and 2, thereby indicating that this rating can partly reflect the disruptiveness of a car.

In model 1, the unweighted lost gene rate was negative, while in model 2, the unweighted new gene rate was negative, and in model 3, the unweighted inherited gene rate was positive.

These coefficients imply that inheriting technology from predecessors is more conducive to enhancing product disruptiveness than introducing new technologies. In other words, at the quantitative level of technology genes, 'small changes' are beneficial for increasing disruptiveness. Conversely, the inherited gene rate difference was negatively associated with the PDI in model 6, the lost gene rate difference was positively related to the PDI in model 4, and the new gene rate difference was negatively related to the PDI in model 5. This suggests that for lost genes, discarding significant genes is advantageous for disruptiveness. For inherited genes, inheriting non-significant genes is favorable for disruptiveness. For new genes, introducing crucial new genes is beneficial for disruptiveness. In other words, altering key technologies within a product is more conducive to enhancing disruptiveness than changing non-essential technologies. At the qualitative level of technology genes, 'not the least changes' contribute to increasing disruptiveness.

In summary, the SBNL (Small But not least Changes) principle facilitates innovative products to better stand on the shoulders of giants and become more disruptive. Building upon the foundation of disruptive ancestors, making minor alterations to key technologies, and maintaining the price within an acceptable range is pivotal for augmenting a product's disruptiveness. The SBNL principle resonates with the MAYA (Most Advanced, Yet Acceptable) principle proposed by Silvennoinen and Mononen (2023), which emphasizes the importance of making the most critical enhancements to a product within a reasonable scope, rather than the most numerous changes.

3.8 Robustness check

To mitigate the potential fixed effects of time and brand on the product's 1-year PDI, we implemented multidimensional regression analyses that account for the fixed effects of both year and brand to ensure robustness. Additionally, we conducted regressions on product gene data computed not only with frequency weight but also with TF-IDF weight, reinforcing the credibility of our results. In Table 6, within the fixed effects regression framework, models 1, 2, and 3 utilize frequency weight, whereas models 4, 5, and 6 leverage TF-IDF weight. TF-IDF is a statistical measure employed to transform a document into a vector, taking into account the significance of the words within the document. The word weights obtained via the TF-IDF method signify the keywords that epitomize the document (Salton and McGill, 1983). For the computation of TF-IDF weights, we treat each year as an individual document. This method is widely applied in phylogenetic networks to evaluate the significance of technologies (Lee et al., 2022; Jeong et al., 2023).

In this section, we directly applied both weighted and unweighted gene rates as explanatory variables, rather than using the indirect variable of gene rates difference. The results demonstrate a notable consistency across most explanatory variables when comparing frequency weight and TF-IDF weight groups. The positive relationship between Ancestor PDI and 1-year PDI suggests that building upon the groundwork laid by giants facilitates the creation of disruptive products. Conversely, there is a negative correlation between product price and PDI, indicating that disruptive products are likely to emerge from the lower end of the market or possess a price advantage (Christensen, 2000; Christensen and Raynor, 2003; Govindarajan and Kopalle, 2006; Tellis, 2006). A negative brand-new gene rate suggests that introducing new technologies, as opposed to capitalizing on established key technologies, does not markedly increase disruptiveness when absorbing the year and brand fixed effects. Indeed, an overabundance of brand-new genes might actually reduce a product's disruptiveness, since the adoption of too many untested, market-unrecognized brand-new genes does not necessarily lead to an increase in technology adopters. However, as evidenced by Table 2, when the fixed effects of year and brand are not accounted for, the brand-new gene rate appears to be conducive to disruptiveness. This also indicates that there is heterogeneity present within the years and brands. The positive unweighted inherited gene rate, contrasted with the negative weighted inherited gene rate, indicates that inheriting predecessor technologies benefit product disruptiveness, however it requires modification of key technologies to introduce greater disruptiveness. A positive weighted lost gene rate versus a negative unweighted lost gene rate implies that discarding significant rather than inconsequential technologies yields more disruptiveness. Moreover, a positive weighted new gene rate, as opposed to a negative unweighted new gene rate, reveals that introducing new and significant technologies, rather than ordinary ones, enhances product disruptiveness. Finally, in all the models, the R-squared values exceed 0.5 which is higher than those of pooled OLS models in Table 5, indicating a higher explanatory power of the product gene rates and other information for the 1-year PDI in our models when absorbing the year and brand fixed effects.

Table 6. Fixed effects regression results from years 2014 to 2023

|  | Frequency weight | | | TF-IDF weight | | |
| --- | --- | --- | --- | --- | --- | --- |
|  | (1) | (2) | (3) | (4) | (5) | (6) |
| Unweighted inherited gene rate | 0.247*** |  |  | 0.543*** |  |  |
|  | (0.046) |  |  | (0.047) |  |  |
| Weighted inherited gene rate | -0.209* |  |  | -0.383*** |  |  |
|  | (0.098) |  |  | (0.038) |  |  |
| Ancestor PDI | 0.581*** | 0.581*** | 0.581*** | 0.521*** | 0.524*** | 0.521*** |
|  | (0.024) | (0.024) | (0.024) | (0.025) | (0.024) | (0.025) |
| Review rate | 0.009+ | 0.009+ | 0.009+ | 0.010+ | 0.008 | 0.010+ |
|  | (0.005) | (0.005) | (0.005) | (0.005) | (0.005) | (0.004) |
| Price(log) | -0.013* | -0.013* | -0.013* | -0.013* | -0.012* | -0.013* |
|  | (0.006) | (0.005) | (0.006) | (0.005) | (0.005) | (0.005) |
| Brand-new gene rate | -0.215*** | -0.235*** | -0.215*** | -0.432*** | -0.499*** | -0.432*** |
|  | (0.046) | (0.044) | (0.046) | (0.047) | (0.055) | (0.047) |
| Weighted lost gene rate |  | -0.020 |  |  | 0.456*** |  |
|  |  | (0.126) |  |  | (0.046) |  |
| Unweighted lost gene rate |  | -0.178*** |  |  | -0.622*** |  |
|  |  | (0.052) |  |  | (0.055) |  |
| Weighted new gene rate |  |  | 0.209* |  |  | 0.383*** |
|  |  |  | (0.098) |  |  | (0.038) |
| Unweighted new gene rate |  |  | -0.247*** |  |  | -0.543*** |
|  |  |  | (0.046) |  |  | (0.047) |
| Year fixed effects | Yes | Yes | Yes | Yes | Yes | Yes |
| Brand fixed effects | Yes | Yes | Yes | Yes | Yes | Yes |
| N | 3694 | 3694 | 3694 | 3694 | 3694 | 3694 |
| R2 | 0.529 | 0.528 | 0.529 | 0.553 | 0.556 | 0.553 |

*Notes:* This table evaluates the relationship between different measures of the use of product genetic information, survey data from website, price and 1-year PDI. Estimates are from ordinary-least-squares regressions. Each coefficient is tested against the null hypothesis of being equal to 0 using a two-sided t-test. We do not adjust for multiple hypothesis testing. Robust standard errors are shown in parentheses.

+ p<0.1; * p<0.05; ** p<0.01; *** p<0.001.

## 5. Conclusion

    Most scholars regard disruptive innovation as a market phenomenon, primarily because its criterion for disruption is predicated on usurping the market share of incumbents. Moreover, it subverts incumbent technologies by belonging to a distinct technological paradigm (Christensen et al., 2004; Christensen and Raynor, 2003). However, from a technological standpoint, although the lean approach (Brad et al., 2016; Tuli et al., 2015), the Minimum Viable Product (MVP) concept (Olsen, 2015; Di Guardo et al., 2022), and the S-curve model (Borgianni and Rotini, 2012; Bradley and O'Toole, 2016) have played certain roles in the design of disruptive products, there is a paucity of literature elucidating the design principles for creating disruptive products based on existing product information. This gap stems from the predominantly qualitative nature of prior research on disruptive innovation (Schmidt and van der Sijde, 2022; Nagy et al., 2016), which has rendered the quantification of product disruptiveness challenging. The PDI method proposed in this paper quantifies the disruptiveness of products using comprehensive technical information available in the market. The PDI is a flexible application of CD index within the product domain, integrating the concept of the product phylogenetic network. The CD index quantifies the disruptiveness of various technologies and innovations through a citation network (Funk and Owen-Smith, 2017). This study employs the PDI method in an empirical study of the passenger car market's product data. Although the approach is technical, the product data indirectly reflects market demand, as products that do not cater to the market are phased out, even though the data does not reveal the specific reasons for their elimination. The contributions of this study are delineated into three key points. Firstly, we have devised the PDI, an innovative methodology for quantifying the disruptiveness of products. Secondly, our research into automotive products has revealed the design principles of SBNL. Additionally, we have observed the phenomenon of the disruptive ancestral effect.

    Firstly, the PDI is a method for measuring the disruptiveness of products. Its theoretical foundation is based on the CD index method, which measures the disruptiveness of innovation through citation networks (Funk and Owen-Smith, 2017), and the product phylogenetic network, which is grounded in the theory of product evolution (Lee et al., 2022). We have employed the concept of the product family triangle to transform the product phylogenetic network into a citation network. The CD index has been validated as a measure capable of quantifying the disruptiveness of innovation (Funk and Owen-Smith, 2017). Furthermore, the observation that the average disruptiveness of innovation is on an annual decline has been substantiated by Park et al. (2023). This evidence serves to corroborate the validity of the PDI. Both the average PDI across all products and the average normalized PDI within product series demonstrate a consistent year-over-year reduction, mirroring the overarching trend of diminishing innovation disruptiveness over time. In addition, we conducted two case studies. Initially, we examined the 2014 Tesla Model S and the 2014 Chevrolet Spark EV, both electric vehicles. Despite their review rates surpassing 4.5 in 2014, our PDI analysis indicates that Tesla's PDI is markedly higher than that of the Chevrolet Spark. We also compared luxury and regular cars from three emblematic automotive companies which are Toyota Motor Corporation, Hyundai Motor Group, and Volkswagen AG. The

PDI findings suggest that the regular cars' PDI from the same manufacturer exceeds that of the luxury cars. This finding is in line with the theory that disruptive innovation typically emerges from the lower end of the market (Christensen, 2000; Christensen and Raynor, 2003; Govindarajan and Kopalle, 2006; Tellis, 2006).

Secondly, we conducted a regression analysis on 3,694 car models from 2014 to 2023. The independent variable was the PDI, while the explanatory variables were three types of gene rates: inherited, lost, and new gene rate. By comparing the coefficients of weighted and unweighted gene rates in the regression from Table 6, as well as analyzing the gene rate differences coefficients in Table 5, we deduced that the principles of SBNL are crucial for enhancing a product's disruptiveness. Specifically, unweighted gene rates represent a quantitative comparison of genes between the focal product and its ancestors, whereas weighted gene rates compare the quality of genes, assigning greater weight to more important technologies. Contrary to intuitive belief, a positive coefficient for the unweighted inherited gene rate suggests that inheriting more genes from ancestors can actually benefit the disruptiveness of the product, which constitutes 'small changes.' Conversely, a negative coefficient for the weighted inherited gene rate indicates that altering key technologies is necessary to achieve greater disruptiveness, which entails 'not least changes.' These are the design principles of SBNL.

Thirdly, our regression model reveals that the Ancestor PDI is consistently and significantly positive in both Table 5 and Table 6. We refer to this phenomenon as the 'disruptive ancestral effect.' It suggests that by inheriting traits from more disruptive ancestors, a focal product may achieve greater disruptiveness compared to other products. Although this concept has not been explicitly identified in other literature, it resonates with Newton's famous assertion: 'If I have seen further than others, it is by standing upon the shoulders of giants.' While the specific mechanisms behind the disruptive ancestral effect remain unclear, it is acknowledged that disruptive innovation can create new markets (Bower and Christensen, 1995; Rhéaume and Gardoni, 2017). Following a highly disruptive ancestor in the exploration of new market spaces increases the likelihood of attaining higher disruptiveness. Disruptive innovation is not about large differences from its predecessors but about standing better on the shoulders of giants.

In addition to addressing the academic gap identified in previous literature—where existing theories of disruptive innovation suggest that disruptive products emerge from lower-end markets and gradually overtake established products (Christensen, 2000; Christensen and Raynor, 2003; Govindarajan and Kopalle, 2006; Tellis, 2006) but do not explain how to design such products by learning and recombining technologies from previous products—this study provides a technical approach to measuring product disruptiveness and derives the SBNL design principles through the analysis of automotive data. Products do not emerge in a vacuum; they result from assimilating existing technologies from predecessor products, incorporating new technologies, and recombining them (Arthur, 2007). SBNL provides a methodological approach to selecting existing technologies and integrating new ones during the product design process. It is not merely a set of design principles but also a means to calculate various gene rates to determine a product's PDI. This serves as a quantitative metric for product design, aiding companies in enhancing their design capabilities. Moreover, for business managers, choosing a product strategy is crucial, as a follower strategy often yields a late-mover advantage (Shankar et al., 1998; Querbes and Frenken, 2017). The disruptive ancestral effect offers a methodology for selecting the correct entities to follow. By referencing the PDI, one can choose an appropriate predecessor to emulate.

In our study, we calculated a 1-year PDI to reflect the short-term disruptiveness of car products, given the rapid update and iteration cycle in the automotive industry, where new models are introduced annually for each series. Consequently, there is no necessity to compute a long-term PDI, which represents a limitation of our research. For other products, a more extended PDI could be calculated. Additionally, when computing various gene rates, we focused on the three ancestors most similar to the focal product. The number of ancestors can be selected based on the researcher's needs. Furthermore, alternative methods could be employed when converting the phylogenetic network into a citation network. Future research could integrate market data with the technical data of the PDI for a more comprehensive study of disruption. The PDI is also suitable for scenarios where market data is challenging to obtain, but technical data is readily accessible. For instance, it is difficult to acquire sales data for all companies' AI products, yet technical data for AI products can be easily obtained from public sources. In addition, due to the existence of uncited ancestors, the CD index will be overestimated. We can use the PDI method to reconstruct the citation network for patents and papers and calculate a more realistic disruptiveness.

Data Availability Statement:
The data that support the findings of this study are available on request from the corresponding author. The data are not publicly available due to privacy or ethical restrictions.

Disclosure Statement: The authors declare no conflict of interest.

Acknowledgement
This research was supported by the BK21 Fostering Outstanding Universities for Research (FOUR) funded by the Ministry of Education (MOE, Korea) and a National Research Foundation of Korea (NRF) grant funded by the Korean government (MSIT) under Grant 2022R1A2C1091917. We appreciate the support and cooperation of the Edmunds.com, which provided valuable car data for this study and authorized us to use them.